\documentclass[aps,pre,twocolumn,groupedaddress,amssymb]{revtex4}
\usepackage{graphicx}
\usepackage[figuresright]{rotating}
\begin{document}

\def\mdr{\mathit{\delta r}}
\def\tmod{|\tau|}
\def\ttod{|t|}

\title{Critical amplitude ratios of the Baxter-Wu model}

\author{Lev N. Shchur$^{1,a}$ and Wolfhard Janke$^{2,b}$}
\affiliation{${^{1}}$Landau Institute for Theoretical Physics,
142432 Chernogolovka, Russia \\
${^{2}}$Institut f\"ur Theoretische Physik and Centre for
Theoretical Sciences (NTZ), Universit\"at Leipzig,
Postfach 100\,920, 04009 Leipzig, Germany \\
\sl e-mail: $^{a}$lev@landau.ac.ru;
$^{b}$Wolfhard.Janke@itp.uni-leipzig.de}

\begin{abstract}

A Monte Carlo simulation study of the critical and off-critical
behavior of the Baxter-Wu model, which belongs to the universality
class of the 4-state Potts model, was performed. We estimate the
critical temperature window using known analytical results for the
specific heat and magnetization. This helps us to extract reliable
values of universal combinations of critical amplitudes with
reasonable accuracy. Comparisons with approximate analytical
predictions and other numerical results are discussed.
\end{abstract}

\maketitle

\section{Introduction}
\label{sec-intro}

One of the central results of the theory of phase transitions and
critical phenomena is the formulation of the universality
hypothesis~\cite{U1,U2}. According to the theory, all systems with
the same dimensionality, the same symmetry of the ordered phase and
the same number of order parameters are described by the same set of
critical exponents at the critical point. Additionally, thermodynamic
functions vary with temperature in such a way that some
combinations of their amplitudes take the same values for all
systems within a universality class~\cite{Priv-rev}. For many systems,
critical exponents are known by  approximate methods (field theoretical
perturbation theory, series expansions, Monte Carlo simulations) and
have been derived exactly in some cases, mainly for two-dimensional
models~\cite{McCoy-Wu,denNijs79,Pearson80,Nienhuis84,Dotsenko84,DotsenkoFateev84}.
The extensive research in the last 50 years gives strong support for
the universality hypothesis in the context of the critical
exponents~\footnote{We have to note some recent discussion of the
weak universality of correlation length amplitudes~\cite{Chen-Dohm,Dohm}
and associated quantities like Binder cumulants~\cite{SS,Selke} in the
case of anisotropic models.}. At the same time, the issue of
critical amplitude ratios was checked only for a few models~\cite{DM-rev}.

Special interest in the properties of universality classes derives from
cases in which the singular behavior is complicated by logarithmic
corrections \cite{log-corr}. There are some systems which belong to the same
universality class but whose off-critical behavior
may be modified by logarithmic corrections. This is the case for
the universality class of the two-dimensional 4-state Potts model.
The model, which gives the name to this universality class, the 4-state
Potts model~\cite{Potts} contains logarithmic corrections to the
critical behavior~\cite{NS,CNS,Salas97,Shchur09} of thermodynamic quantities.
For example, the free energy $F$ and the magnetization $M$ in the
ordered phase behave at criticality as~\cite{NS,CNS}
\begin{eqnarray}
F(\tau)&\approx& A_{\rm 4p} \tmod^{2-\alpha} [\ln(-\tau)]^{\alpha_l} \,, \label{F-4-potts} \\
M(-\tau)&\approx & B_{\rm 4p} (-\tau)^\beta [\ln(-\tau)]^{\beta_l} \,,
\label{mag-4-potts}
\end{eqnarray}
where $\alpha{=}2/3$, $\beta{=}1/12$, $\alpha_l{=}{-}1$,
$\beta_l{=}{-}1/8$, and $\tau{=}1{-}T_c/T$ is a measure of the
distance of the temperature $T$ to the critical temperature $T_c$.

Another two-dimensional model, which belongs to the universality
class of the 4-state Potts model, is the Baxter-Wu model \cite{BW}. For this
model it is known exactly~\cite{BW,BW-aus,Baxter-aus,Baxter} that
close to the critical temperature, the free energy and the
magnetization in the ordered phase behave as
\begin{eqnarray}
F(\tau)&\approx& A_{\rm bw} \tmod^{2-\alpha} \,,\\
M(-\tau)&\approx& B_{\rm bw} (-\tau)^\beta \,, \label{mag-bw}
\end{eqnarray}
with the same values for the critical exponents as in (\ref{F-4-potts})
and (\ref{mag-4-potts}), but without logarithmic
corrections to the singular behavior.

A possible explanation for the difference in the off-critical
behavior of the Baxter-Wu and 4-state Potts models is that for some unknown
reason the coefficient behind the logarithmic correction is zero for the
Baxter-Wu model~\cite{NS,CNS}, although the leading critical behavior
for both models seems to be described by the same fixed point in the
renormalization-group space~\cite{KDA}.

Critical amplitudes are not universal and one should not expect that
the free-energy amplitudes $A_{\rm bw}$ of the Baxter-Wu model and
$A_{\rm 4p}$ of the 4-state Potts model are equal. At the same time,
some combinations of the amplitudes are universal. For example, the
ratio of the free-energy amplitude $A_{\rm bw}(+)$ in the high-temperature
phase of the Baxter-Wu model to the amplitude $A_{\rm bw}(-)$ in the
low-temperature phase equals unity, as well as the ratio of the corresponding
amplitudes of the 4-state Potts model $A_{\rm 4p}(+)/A_{\rm 4p}(-){=}1$.
This is a consequence of the duality relation for the free energy of
the two models~\cite{Baxter}.

The ratio of the susceptibility amplitudes ${\cal
R}_\chi{=}\Gamma_{\rm 4p}(+)/\Gamma_{\rm 4p}(-){=}\Gamma_{\rm bw}(+)/\Gamma_{\rm bw}(-)$,
on the other hand, is not known exactly and different approximations (analytical,
series expansions, Monte Carlo data) from different groups are not
coherent and differ substantially (for a recent discussion, see
the papers~\cite{Shchur09,bbjs}). It is well known that logarithmic
corrections, if they exist, complicate the critical behavior and render
the analysis of the critical behavior extremely difficult if possible at all, and
the determination of critical exponents from Monte Carlo (MC) and series
expansions (SE) became non-trivial. In fact, the determination of critical
exponents and corrections to scaling may lead to indecisive
conclusions~\cite{DelfinoBarkemaCardy00}.

The purpose of the present paper is to estimate numerically critical
amplitudes of the Baxter-Wu model, which is free of logarithmic corrections,
and to compare universal amplitude ratios with the ones available for the
4-state Potts model. It should be
emphasized that critical amplitude ratio universality is a non-local
property of the renormalization group phase space whereas the critical
exponents characterize its behavior only in the vicinity of
the corresponding fixed point.

We use the traditional Metropolis MC algorithm to
simulate the Baxter-Wu model, and analyze the magnetization and
polarization in the ordered phase, and the energy, specific heat
and magnetic susceptibility in both phases. We estimate the accuracy of our
data by comparing the magnetization, polarization, energy and specific
heat to available exact results.

In particular, we use the known results for the energy, specific
heat, magnetization and polarization in order to estimate the
critical temperature window, i.e., the range of temperatures in
which the system on a finite lattice behaves to a very good
approximation as on an infinite one. This allows us to estimate the
critical ratio of the susceptibility amplitudes in the high- and
low-temperature phase with good accuracy, ${\cal R}_{\chi}=3.9\pm
0.1$. The analytical estimate of this ratio, obtained by Delfino and
Cardy~\cite{Delfino98} using some approximation of the exact
scattering field theory~\cite{ChiZ92}, is only slightly larger, ${\cal
R}_{\chi}=4.013$. Delfino and Grinza~\cite{DG04} obtained practically the same
value ${\cal R}_{\chi}=4.02$ using the same approximation for the Ashkin-Teller model with parameters
which correspond to the 4-state Potts model universality class.
A recent analysis of the
MC and SE data for the 4-state Potts model gives an amplitude ratio in
the range of about $6.5(4)$ ~\cite{SBB-letter,bbjs,Shchur09}.
At the same time, the values for the universal combination of amplitudes
$R_C^-=\alpha A_0 \Gamma_-/B_0^2$ in the low-temperature phase reported in
Ref.~\cite{Delfino98} and Refs.~\cite{SBB-letter,bbjs,Shchur09} are
$0.00508$ and $0.0052(2)$, respectively ($A_0$ and $B_0$ are the specific-heat and
magnetization amplitudes). This is in perfect agreement with our estimate $0.00517(7)$
we present here for the Baxter-Wu model.

In the rest of the paper we present the details of our analysis of the amplitude
ratios for the Baxter-Wu model. In Section~\ref{sec-model} and
\ref{sec-previous} we first give an overview of known analytical results
and previous numerical simulations which support them. We then discuss in
Section~\ref{sec-simulations} details of our simulation algorithm
realization, including the special choice of the lattice, averaging,
etc. Section~\ref{sec-results} presents the details of our critical
amplitude estimation, and the discussion in Section~\ref{sec-discussion}
summarizes our results and touches on some open questions.

\section{Model and exact results}
\label{sec-model}

In this section we summarize those known analytical results for the
Baxter-Wu model which will be used in the lattice construction,
algorithm realization and data analysis.

\subsection{Model}

The Baxter-Wu model is defined on a triangular lattice, with spins $\sigma_i=\pm
1$ located at the vertices. The three spins forming a triangular face
are coupled with a strength $J$, and the Hamiltonian reads

\begin{equation}
{\cal H} =-J\sum_{\rm faces} \sigma_i\sigma_j\sigma_k \,,
\label{ham-bw}
\end{equation}

\noindent where the summation extends over all triangular faces of the
lattice, both pointing up and down.

\subsection{Self-duality and critical temperature}

The model is self-dual as found by Wood and Griffiths~\cite{Wood72}
and Merlini and Gruber~\cite{Merl-Grub}, who applied the Kramers-Wannier
construction developed for the square-lattice Ising model, and showed
that both models, Ising on the square lattice and Baxter-Wu on
the triangular lattice, share the same self-dual temperature (see also the book~\cite{Baxter}).
\noindent The partition function reads
\begin{equation}
Z=\sum_{\sigma} \exp \left[\beta\sum_{\rm faces}
\sigma_i\sigma_j\sigma_k\right] \,, \label{part-Z}
\end{equation}

\noindent where $\beta=J/k_BT$. The dimensionless free energy per
site is
\begin{equation}
f/k_BT \equiv \psi(\beta)=-\lim_{N\to\infty}N^{-1}\ln Z \,,
\label{free-en}
\end{equation}

\noindent where $N$ is the number of lattice sites.
It satisfies the duality relation

\begin{equation}
\psi(\beta)=\psi(\beta^*)+\ln\left(\sinh2\beta^*\right) \,,
\label{dual-bw}
\end{equation}

\noindent where
\begin{equation}
\sinh 2\beta^* \sinh 2\beta=1 \,. \label{dual-beta}
\end{equation}

\noindent This is precisely the duality relation of the square
lattice Ising model. The argument of Kramers-Wannier can be applied:
if there exists just one  critical point, then it must occur when
$\beta=\beta_c=\beta_c^*$, where
\begin{equation}
\sinh 2\beta_c =1 \,, \;\;\; \beta_c=\ln{(\sqrt{2}+1)}/2 \,. \label{crit-K}
\end{equation}

\subsection{Ground-state symmetry}

Let $\sigma_A$, $\sigma_B$, $\sigma_C$ denote all the spins on the
$A$, $B$, $C$ sub-lattices, respectively. Any total
configuration $(\sigma_A,\sigma_B,\sigma_C)$ of spins has the
same energy as three others. These configurations can be obtained by flipping all
spins on two of the sub-lattices. The spin configurations can be grouped in
equal-energy  sets of four:
\begin{eqnarray}
(\sigma_A,\sigma_B,\sigma_C)&,&(\sigma_A,-\sigma_B,-\sigma_C) \,, \nonumber \\
(-\sigma_A,\sigma_B,-\sigma_C)&,&(-\sigma_A,-\sigma_B,\sigma_C) \,.
\label{conf-four}
\end{eqnarray}
The ground state is thus four-fold degenerate: one ferromagnetic
state with magnetization (per site) $m=1$ and three ferrimagnetic
states with $m=1/3$.

\subsection{Exact solution and critical behavior}

Baxter and Wu solved the model at the critical temperature exactly~\cite{BW,BW-aus,Baxter-aus,Baxter}
and found that the critical value of the energy
$e = \langle {\cal H} \rangle/N$ is $e_0=-\sqrt{2}|J|$,
the specific heat $C = de/dT$ diverges at $T_c$ as~\cite{BW,BW-aus}
\begin{equation}
C \propto |t|^{-2/3} \,,
\label{spec-heat-bw}
\end{equation}
and the critical behavior of the magnetization for $\tau \le 0$ is~\cite{LT-exact}
\begin{equation}
m \propto |t|^{1/12} \,,
\label{bw-mag-crit}
\end{equation}
where here the reduced temperature is defined as $t = (T-T_c)/T_c$.
The critical
exponents thus take the values $\alpha{=}2/3$ and $\beta{=}1/12$, which
is the two-dimensional 4-state Potts model universality class.

\subsection{Joyce's results for $C$ and $M$}

Joyce established analytic properties of the free energy per spin and
found the explicit form~\cite{Joyce-C}
\begin{equation}
-\frac{f}{k_BT}=\frac{2|J|}{k_BT}+\ln \Lambda(u) \,,
\label{free-expl-joyce}
\end{equation}
where
\begin{equation}
\frac{1}{\Lambda(u)}=\frac{1}{(1+u)^2}
{_2F_1}\left[\frac12,\frac16;\frac43;\frac{16u(1-u)^2}{(1+u)^4} \right] \,,
\label{Lambda-joyce}
\end{equation}
with ${_2F_1}$ denoting the hypergeometric function and
$u=\exp(-4|J|/k_BT)$ such that the critical point value in this variable
is $u_c=3-2\sqrt{2}$.

The specific heat critical behavior follows as
\begin{equation}
\frac{C(t)}{k_B}=A_0\ttod^{-\frac23}+A_1+A_2t \ttod^{-\frac23}+A_3\ttod^{\frac23}+{\cal O}(t) \,,
\label{heat-joyce}
\end{equation}
where
\begin{eqnarray}
A_0&=&\frac29
\left(\ln(\sqrt{2}+1)\right)^\frac43=0.187\,787\,867\ldots\,
, \\
A_1&=&-\frac12 \left( \ln(\sqrt{2}+1)\right)^2=-0.388\,409\,700 \ldots\,
, \\
A_2&=&\frac{2}{27} \left( \ln(\sqrt{2}+1)\right)^\frac43
\left( 7\sqrt{2}\ln(\sqrt{2}+1)-4\right) \nonumber \\
&=& 0.295\,775\,490 \ldots\, , \\
A_3&=&\frac59 \left(\ln(\sqrt{2}+1)\right)^\frac83=0.396\,723\,182\ldots\,
.
\label{heat-joyce-coeff}
\end{eqnarray}

The magnetization reads~\cite{Joyce-M} ($t \le 0$)
\begin{equation}
m=\ttod^{\frac{1}{12}}\left(B_0+B_1\ttod^{\frac23}+B_2\ttod+B_3\ttod^{\frac43}+
{\cal O}(\ttod^{\frac53})\right) \,,
\label{M-joyce-t}
\end{equation}
where
\begin{eqnarray}
B_0&=&2^\frac38 \left(\ln(\sqrt{2}+1)\right)^\frac{1}{12}=1.283\,264\,709\ldots \,,\\
B_1&=&-\frac{1}{2^\frac58}\left(\ln(\sqrt{2}+1)\right)^\frac34=-0.589\,829\,210\ldots \,,
\\
B_2&=&\frac{1}{24\cdot 2^\frac58}\left(\ln(\sqrt{2}+1)\right)^\frac{1}{12}
\left(4-\sqrt{2}\ln(\sqrt{2}+1)\right) \nonumber \\
&=&0.073\,615\,269\ldots \,, \\
B_3&=&\frac{1}{2^\frac58}\left(\ln(\sqrt{2}+1)\right)^\frac{17}{12}=0.542\,208\,469\,6\ldots
\,.
\label{mag-expansion}
\end{eqnarray}

Some time later, Baxter~\cite{Baxter} found an elegant form for the free
energy in terms of an infinite sum of the ratio of some polynomials.
The magnetization (and also the polarization defined below in Eqs.~(\ref{pol-d}) and~(\ref{bw-pol}))
can be expressed in terms of an infinite product
of ratios of polynomials. These expressions may be used to obtain
the energy, specific heat, magnetization and polarization with any
desired accuracy. The summary of analytical results is presented in this
section in a form which is most suitable for the analysis of the data
discussed here.

\section{Previous numerical results}
\label{sec-previous}

\subsection{Monte Carlo study of the critical behavior of the Baxter-Wu model}

In course of their simulation study of the influence of quenched
impurities on the critical behavior, Novotny and
Landau~\cite{Mark-David81} performed for comparison also first Monte
Carlo analyses of the pure Baxter-Wu model. They defined the
magnetic order parameter $m$ as the root-mean-square (rms)
of the magnetization on the three sub-lattices. The polarization
order parameter $p$ was defined as the rms average of the two-spin
correlation functions for the nearest-neighbor spins between
different sub-lattices. They pointed out that the rms
rather
than the
the absolute value of the sum of the sub-lattice
magnetizations was used because the susceptibilities calculated using
the rms
definition
gave results that were in closer agreement with
high- and low-temperature series-expansion results.

Finite-size scaling plots of the order parameters, the magnetization
$mL^{\beta_m/\nu}$ and polarization $pL^{\beta_p/\nu}$, as functions
of $t L^{1/\nu}$ are consistent with $\beta_m{=}\beta_p{=}\beta{=}1/12$ and
$\nu{=}2/3$. In the symmetric phase ($T>T_c$) those plots show in a
log-log representation a decay with slope $-7/12$, approaching asymptotic
values at very large values of $t L^{1/\nu}$. The energy, specific heat,
and susceptibility also behave according to the Baxter-Wu predictions.

They simulated lattices with linear size $L$ up to $66$
and estimated magnetic susceptibility amplitudes,
$\Gamma_{\rm bw}(+){=}0.03(2)$ and
$\Gamma_{\rm bw}(-){=}0.010(5)$, and polarization amplitudes,
$\Gamma_{\rm bw}^p(+){=}0.06(3)$ and
$\Gamma_{\rm bw}^p(-){=}0.04(2)$. Therefore, the ratio of amplitudes
is about $1.5-3$ with an uncertainty of about 50 per cent.

\subsection{Monte Carlo renormalization-group study of the Baxter-Wu
model}

In their Monte Carlo renormalization-group (MCRG) analysis, Novotny,
Landau, and Swendsen~\cite{NLS82} did not find any evidence for
logarithmic corrections in the Baxter-Wu model on lattices with
sizes $21\times 21$, $42\times 42$, and $147\times 147$. For the RG
eigenvalues they estimated $y_T{=}1.48\pm 0.03$, $y_h{=}1.875\pm 0.003$,
and results for $y_3$ were consistent with the Barber
ansatz~\cite{Barber76} $y_3=7/8$, based on the renormalization group
analysis. The latter exponent is the correction-to-scaling magnetic
exponent $y_{\sigma_2}=2-x_{\sigma_2}$
(see, Refs.~\cite{DotsenkoFateev84,Nienhuis84} and the discussion
in Ref.~\cite{bbjs}), which
produces the correction-to-scaling exponent $2/3$ in the
specific heat~(\ref{heat-joyce}) and magnetization~(\ref{M-joyce-t}).

\subsection{Conformal invariance studies of the Baxter-Wu model
and a related site-coloring problem}

The operator content of the Baxter-Wu model has been discussed by Alcaraz and
Xavier~\cite{AX-cft,AX-operator}, who extended the original Bethe
ansatz solution of the site-coloring problem and solved numerically
the corresponding equations of the transfer matrix for toroidal boundary
conditions.
They found that the latter problem has the same operator content as the
4-state Potts model. The correction-to-scaling effects seem,
however, to correspond to different perturbations of the fixed point
of the renormalization group. The authors of Refs.~\cite{AX-cft,AX-operator}
claim that the correction-to-scaling terms contain only integer powers,
like those in the Ising model.

\section{Simulations}
\label{sec-simulations}

\subsection{Lattice choice}

Genuine critical behavior can be observed only in the thermodynamic limit,
when the system approaches infinite size. Simulations, on the other hand,
are always performed on finite lattices~\footnote{At
least in the directions of the ``space'' dimensions, as is the case for the
transfer-matrix method where the ``time'' direction can be taken to infinity.}.
Finiteness of the lattice leads to such
effects as scaling of thermodynamic quantities with the lattice
size at criticality~\cite{Barber-v8}, and the shift~\cite{DPLandau76} of
the pseudo-critical point~\footnote{The temperature
at which thermodynamic quantities, such as the specific heat and susceptibility,
reach a maximum value. The thus defined pseudo-critical temperatures reach $T_c$
in the limit of infinite lattice size, and the maximum value of thermodynamic
quantities diverges accordingly.}.

An additional source for systematic deviations in the simulations is due to the
approximation of the aspect ratio which is often overlooked.
The central idea is to choose such a form
of the finite lattice and boundary conditions for which the number of sites and number of
bonds would be the same in either direction, and on both the prime
and its dual lattice~\footnote{It has been demonstrated that such
implementation is very helpful in simulations of the
bond-percolation problem on the square lattice~\cite{KS97}.}.
This is the property of the model on the infinite lattice.

\begin{figure}
\centering
\includegraphics[angle=0,width=\columnwidth]{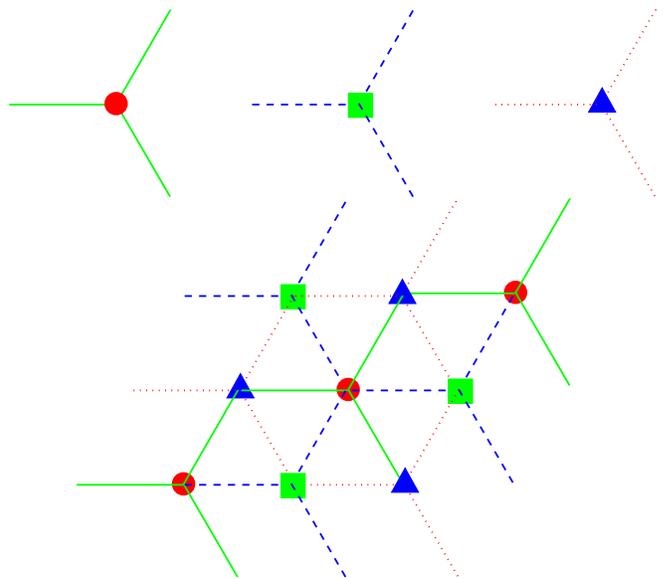}
\caption{(Color online) Top row: Three elements generating the lattice with
sites R, G, and B (circle, box, and triangle) and
bonds G (solid line), B (dashed line), and R (dotted line), attached
correspondingly. Bottom: The smallest lattice having 3 sites, 3 bonds, and
zero sum of site colors and bond colors in any of the three lattice directions
under periodic boundary conditions, $L{=}3$.}
\label{fig-drei-elem}
\end{figure}

We construct the triangular lattice by using the three elements shown in the top row
of Fig.~\ref{fig-drei-elem}. We associate one of the three colors with each site
and bond, and use the convention that the sum of R (red), G (green), and B (blue)
vanishes, or, equivalently, equals to W (white) color.
The rule of construction is that a site of the same color should never be a
neighbor, and the same condition should hold for the bonds.
This is the natural partition of the Baxter-Wu lattice on the three sub-lattices.
We denote in Fig.~\ref{fig-drei-elem} a R-site with a circle, a G-site with a box, and
a B-site with a triangle.  G-bonds are denoted in Fig.~\ref{fig-drei-elem} with a
solid line, B-bonds with a dashed line, and R-bonds with a dotted line.

\begin{figure}
\centering
\includegraphics[angle=0,width=\columnwidth]{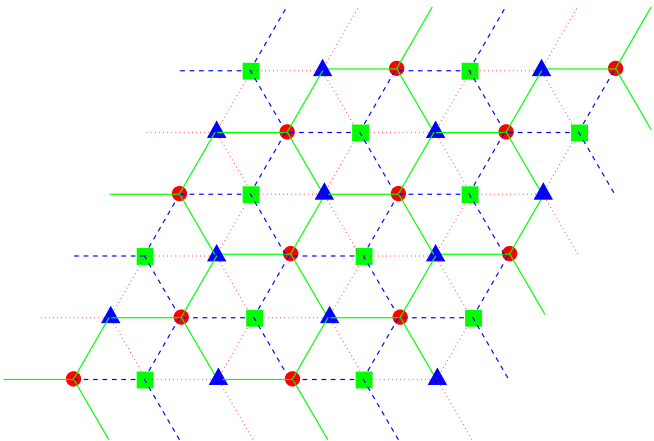}
\caption{(Color online) The next smallest $6\times 6$ lattice with
6 sites, 6 bonds, and ``zero'' or ``white'' colors in any lattice direction
under periodic boundary conditions, $L{=}6$.}
\label{fig-drei}
\end{figure}

The minimal lattice size which follows the above mentioned
requirements is shown at the bottom of Fig.~\ref{fig-drei-elem}.
It contains exactly three sites of different colors and three bonds of different
colors in any of the three lattice directions with periodic boundary conditions.
The sum of site colors and bond colors is white
along any lattice direction (by definition, $R+G+B = W \equiv 0$).
This $3\times 3$ lattice can be used as an elementary building block for constructing
larger lattices. An example of the next size of the
lattice, the $6\times 6$ lattice, is shown in Fig.~\ref{fig-drei}.

There are the three sub-lattices for the Baxter-Wu model labeled R, G, and
B, therefore the total number of sites (and bonds) should be a multiple
of three. The bonds between each two sub-lattices form a hexagonal
lattice: for removed sites R it is composed by the hexagons formed by
the red bonds $L_{\rm GB}$, for removed sites G by
the hexagons formed by the green bonds $L_{\rm RB}$, and for removed
sites B by the hexagons formed by the blue bonds $L_{\rm RG}$.

Such construction keeps the symmetry of the Baxter-Wu model.
In addition, the Baxter-Wu model is self-dual and our choice
of the lattice construction keeps self-duality not only in
the thermodynamic limit, but also for any finite size of the lattice.
This in turn minimizes the possible influence of the approximation
for the aspect ratio~\cite{KS97}.

\subsection{Choice of the algorithm}

It is well known that Monte Carlo simulations of the Baxter-Wu model
experience strong finite-size effects and an application of the traditional
Metropolis algorithm becomes costly due to the critical slowing
down. Novotny and Evertz~\cite{NE-cluster} proposed some time ago
a cluster algorithm for the Baxter-Wu model. Recently, this algorithm has been
extended to the generalized self-dual Baxter-Wu model~\cite{DGHBN}. The main
idea is that the Hamiltonian is invariant under the transformation where all
spins on the two sub-lattices are changed. The algorithm fixes the
spins on one sub-lattice and builds up clusters of correlated spins on the
remaining two sub-lattices (see Fig.~\ref{fig-drei}). It
is, however, not obvious that such clusters which live on the subspace of
possible configurations will percolate at the critical point of the
Baxter-Wu model. Indeed, we found that the percolation point of such
clusters seems to be shifted a little bit to lower temperatures. As a result
this leads to the shift of the curves for some observables.  The same effect was found recently for the behavior
of Fortuin-Kasteleyn clusters in the Z$_4$ spin model~\cite{PSS08}. Despite the
slowing-down problem, we therefore resorted in our simulations to the
traditional Metropolis update algorithm.

\subsection{Metropolis algorithm}

To update the spin configurations with the Metropolis algorithm~\cite{MR2T2}
we calculate the local energy of a spin at position $(j,k)$,
\begin{eqnarray}
e_{j,k}=-\sigma_{j,k}&&\!\!\!\!\!\!\left(\sigma_{j,k-1}\sigma_{j+1,k}+
\sigma_{j+1,k}\sigma_{j+1,k+1} \right. \nonumber \\
\!\!\!&\!\!\!\!\!\!\!\!\!\!\!\!\!+\!\!\!\!\!\!&\sigma_{j+1,k+1}\sigma_{j,k+1}
+\sigma_{j,k+1}\sigma_{j-1,k}  \\
&\!\!\!\!\!\!\!\!\!\!\!\!\!+\!\!\!\!\!\!&\left.\sigma_{j-1,k}\sigma_{j-1,k-1}
+\sigma_{j-1,k-1}\sigma_{j,k-1} \right) \,.\nonumber
\end{eqnarray}

\noindent If $e_{j,k}>0$, we flip spin $\sigma_{j,k}$. Otherwise we
flip the spin only if $\exp(2\beta e_{j,k})$ is not less  than a uniformly
distributed random number $\in (0,1]$.

\subsection{Averaging over the ensemble}

In simulations, the specific heat can be found from

\begin{equation}
C=N\frac{1}{k_B T^2}
\left(\langle e^2 \rangle - \langle e \rangle^2\right) \,,
\label{spec-heat-fluc}
\end{equation}
where the energy per site is calculated as
\begin{equation}
e = -\frac{1}{N} \sum_{j,k} \sigma_{j,k}
\sigma_{j+1,k+1}(\sigma_{j+1,k}+\sigma_{j,k+1})
\label{bw-energy}
\end{equation}
and $N=L^2$ denotes the number of sites.

Similarly, the reduced magnetic susceptibility in the low-temperature phase, $\chi_-$,
can be obtained from
\begin{equation}
k_B T\chi_-=N\left(\langle m^2 \rangle - \langle m \rangle^2\right),
\label{susc-fluc}
\end{equation}
where the magnetization $m$ is computed as the sum of the
magnetization over the three sub-lattices,
\begin{equation}
m=\left|\sum_{i=1}^3 m_i\right| \,,
\label{bw-mag}
\end{equation}
with the magnetization per site, $m_i$, of the sub-lattice $i$ given by
\begin{equation}
m_i=\frac{1}{N} \sum_{L_i} \sigma_{j,k} \,.
\end{equation}

In the analysis, we actually calculated the magnetization in an alternative way as~\cite{Mark-David81}
\begin{equation}
m_s=\sqrt{\sum_{i=1}^3 m_i^2} \,,
\label{bw-mag-alt}
\end{equation}
since it leads to more accurate results as was already mentioned in
Ref.~\cite{Mark-David81} and will be discussed in more detail below.

The magnetic susceptibility in the high-temperature phase is computed as
\begin{equation}
k_B T\chi_+=N \langle m^2 \rangle \,.
\label{susc-ht}
\end{equation}

The polarization $p$ per site follows as the sum of the polarizations
\begin{equation}
p_i=\frac{1}{N} \sum_{L_{jk}} \sigma_{j,k}
(\sigma_{j,k-1}+\sigma_{j+1,k+1}+\sigma_{j-1,k})
\label{pol-d}
\end{equation}
between two sub-lattices $j$ and $k$:
\begin{equation}
p=\left|\frac13 \sum_{i=1}^3 p_i\right| \,.
\label{bw-pol}
\end{equation}

We compute the average of a quantity $A$ (where $A$ is one of
$e$, $e^2$, $e^4$,
$m$, $m^2$, $m^4$,
$m_s$, $m_s^2$, $m_s^4$,
$p$, $p^2$, $p^4$)
as a sum over $N_{\rm av}$ steps,
$$\langle A
\rangle = \frac{1}{N_{\rm av}}\sum_{n=1}^{N_{\rm av}} A_n \,.$$

\noindent Before
averaging, we equilibrate the system with $N_{\rm relax}$ MC steps. Typically,
$N_{\rm relax}=10^5-10^6$ and $N_{\rm av}=10^6-10^7$.

\begin{figure}[t]
\centering
\includegraphics[width=\columnwidth]{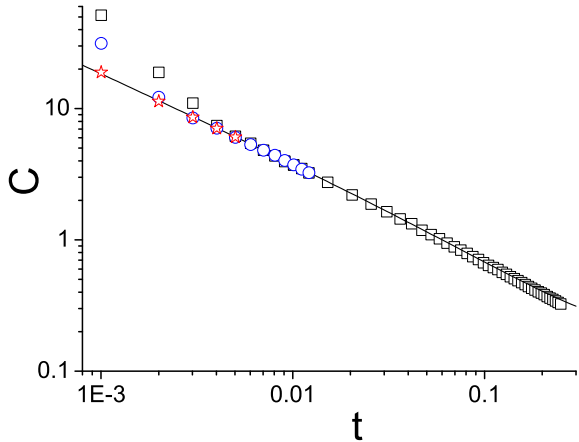}
\caption{Specific heat in the high-temperature phase for system
sizes $L=162$ (open boxes), $L=243$ (open circles), $L=363$ (stars),
and the four-term approximation
Eqs.~(\ref{heat-joyce})--(\ref{heat-joyce-coeff}) to the exact
solution (solid line). Error bars are of the size of the data symbols.} \label{c-ht-fin}
\end{figure}

\begin{figure}[t]
\centering
\includegraphics[width=\columnwidth]{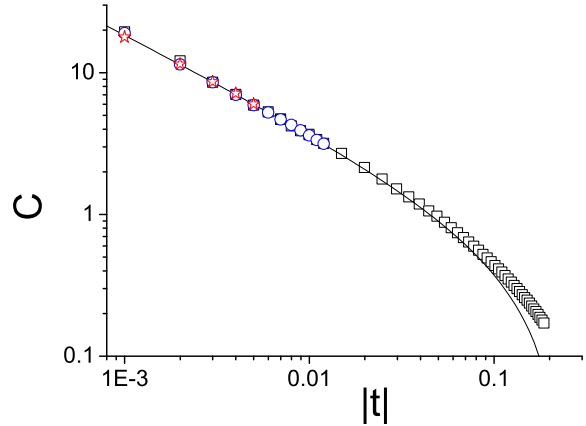}
\caption{Specific heat in the low-temperature phase. Symbols and
curves are the same as in Fig.~\ref{c-ht-fin}.} \label{c-lt-fin}
\end{figure}

\subsection{Dual reduced temperatures}

We compute thermodynamic quantities at the reduced temperature
values $\tau$ and $\tau^*$ connected via the duality
relation~(\ref{dual-beta}) which can be written in the form

\begin{equation}
\tanh\beta^*=e^{-2\beta} \,.
\label{dual-beta-2}
\end{equation}
%
The reduced temperatures $\tau$ and
$\tau^*$ are defined as
\begin{eqnarray}
\tau=\frac{T-T_c}{T}=1 - T_c/T, && \tau>0 \,, \label{dual-t}\\
\tau^*=\frac{T^*-T_c}{T^*}=1 - T_c/T^*, && \tau^*<0 \,.
\label{dual-t-star}
\end{eqnarray}
Due to (\ref{dual-beta-2}) the reduced temperatures are related by
\begin{equation}
\tau=1+\frac{1}{2 \beta_c}\ln\!\left\{\tanh[\beta_c(1-\tau^*)]\right\}, \;
\tau^*<0 \,.
\label{t-relation}
\end{equation}

In some formulas we also employ the reduced temperature $t=\frac{T-T_c}{T_c}$
as in Joyce's papers~\cite{Joyce-C,Joyce-M} and in
Eqs.~(\ref{heat-joyce}), (\ref{M-joyce-t}), which is related to
$\tau$ by $t\approx\tau+\tau^2+{\cal O}(\tau^3)$.

\section{Results}
\label{sec-results}

In this section we employ natural units in which $J = k_B = 1$ and
first define the temperature region window in which we
would fit our data. We use the exact knowledge of the specific-heat behavior
for that purpose. We demonstrate how reliable the fits to the data are.

\subsection{Temperature region window}

The specific-heat data in Figs.~\ref{c-ht-fin} and \ref{c-lt-fin} exhibit
strong finite-size effects close to the critical temperature, that is when the
reduced temperature $t$ approaches zero. This is particularly pronounced in the
high-temperature phase shown in Fig.~\ref{c-ht-fin}, where
one can see visible deviations of the MC data sets from the exact solution
for very small $t$. For temperatures $t > 0.003$, however, the shown data sets
coincide with each other and with the exact result. For that values of
temperature, the correlation length becomes smaller than the system size and
the relation $\xi_0 \propto t^{-\nu} \ll L$ holds better for larger reduced
temperatures.

For large (absolute) values of reduced temperature, the computational data become
more and more close to the exact values. At the same time, the solid line,
which represents an approximation to the exact solution, starts to diverge from
the computational data, because higher-order correction-to-scaling terms,
which are not included in the approximation, become more and more important
for larger reduced temperatures. So, the critical temperature window is bounded for
smaller reduced temperatures $\ttod$ by finite-size effects and for
larger reduced temperatures $\ttod$ by the neglected correction-to-scaling terms
in the analysis.

We have to stress that the left boundary of the temperature window is determined
clearly by the nature of the phase transition -- it is the temperature at which
two length scales coincide: the (randomly) fixed system size and the temperature
dependent correlation length. The right boundary of the temperature window is
not fixed by any physical reason. It depends on the correction-to-scaling
variation with temperature and the number of terms considered in Eq.~(\ref{heat-joyce}).
Here some conventions are necessary. For example, we can define the right boundary
as the temperature up to which the first two terms of the full correction set are
important. Of course, one can also choose one term or three terms. For some
systems or for some particular value of a tuning parameter, it may happen that the
first correction-to-scaling term is close to zero, so that in some wide
temperature region the system would behave as an infinite one, i.e., the
correction-to-scaling terms are not important (see the discussion of such
an extended scaling in Ref.~\cite{Butera-Pernici}).

In the case of the Baxter-Wu model, correction-to-scaling terms are not small.
The power of correction terms decays slowly with the exponent $1/3$ [see
Eq.~(\ref{heat-joyce})]. The amplitudes of correction terms do not depend
on any parameter. So, we have to choose some convention. Using the known
powers and amplitudes of correction-to-scaling terms for both the specific
heat and magnetization, we can estimate the right edge of the temperature
window as that (reduced) temperature for which the relative deviation of
the truncated expansion (\ref{heat-joyce}), denoted by $C_N(\ttod_R)$ with
$N$ terms, from the exact value of $C(\ttod_R)$ is smaller than some
$\epsilon$, $\left| C_N(\ttod_R)/C(\ttod_R)-1\right| < \epsilon$. Fixing
some value of $\epsilon$, say $\epsilon\approx 0.001$ (deviation less than
one tens of per cent), we can estimate the value of $\ttod_R$ as a
function of the number of correction-to-scaling terms $N$ we want to
include in the analysis.

For the data analysis, we will use a combined set of data obtained
for the system size $L=363$ when $\ttod\le 0.02$ and for the size
$L=243$ for larger reduced temperatures $\ttod$, if not mentioned otherwise.
Both data sets are computed with $N_{\rm relax}=10^6$ MC steps for relaxation
and $N_{\rm av}=10^7$ for averaging.

\subsection{Specific heat}

We define effective amplitudes of the specific heat by
\begin{equation}
A_{\rm eff}{(\pm)}(t)=C(t) \ttod^{2/3} \,,
\end{equation}
where the argument ``$\pm$'' refers to the high-temperature
(reduced temperature $t$ is positive) and low-temperature phase
(reduced temperature $t$ is negative), respectively.

\begin{table}[b]
\caption{Results of fits to the MC data for the effective specific-heat
amplitude in the high-temperature phase.}
\center
\begin{tabular}{lllll}
\hline\hline
fit window & $A_0(+)$ & $A_1(+)$ & $A_2(+)$  & $A_3(+)$ \\ \hline
exact (Ref.~\cite{Joyce-C}) & 0.1877$\ldots$ & $-0.3884\ldots$ &
0.2957$\ldots$ & $0.3967\ldots$\\ \hline
$0.001{<}t{<}0.76$   & 0.1878(8) & $-0.42(3)$  & 0.91(7) & $-0.29(5)$ \\ \hline
$0.001{<}t{<}0.32$   & 0.1879(12) & $-0.43(5)$  & 0.93(17) & $-0.31(14)$ \\
                 & 0.1862(8) & $-0.32(2)$  & 0.58(3) & 0 \\ \hline
$0.003{<}t{<}0.76$   & 0.1865(8) & $-0.38(3)$  & 0.82(6) & $-0.24(4)$ \\
                & 0.1827(6) & $-0.24(1)$  & 0.46(1) & 0 \\ \hline
$0.003{<}t{<}0.32$   & 0.1852(12) & $-0.31(6)$  & 0.59(18) & $-0.04(15)$ \\
                & 0.1849(6) & $-0.30(1)$  & 0.54(2) & 0 \\  \hline
$0.001{<}t{<}0.025$   & 0.1893(22) & $-0.52(16)$  & 1.2(5) & 0 \\
                & 0.1846(11) & $-0.16(2)$  & 0 & 0 \\  \hline
$0.001{<}t{<}0.015$   & 0.1860(15) & $-0.20(4)$  & 0 & 0 \\ \hline
$0.001{<}t{<}0.01$   & 0.1870(23) & $-0.24(7)$  & 0 & 0 \\
\hline\hline
\end{tabular}
\vspace*{0.5cm}
\label{tab-c-eff-ht}
\caption{Results of fits to the MC data for the effective specific-heat
amplitude in the low-temperature phase.}
\center
\begin{tabular}{lllll}
\hline\hline
fit window & $A_0(-)$ & $A_1(-)$ & $A_2(-)$ & $A_3(-)$ \\ \hline
exact~(Ref.~\cite{Joyce-C}) & 0.1877$\ldots$ & $-0.3884\ldots$ &
0.2957$\ldots$  & $0.3967\ldots$ \\ \hline
$0.001{<}\ttod{<}0.4$   & 0.1895(6) & $-0.457(28)$ & $0.615(88)$ & 1.05(7) \\
                     & 0.1972(11) & $-0.863(22)$  & $  -0.679(30)      $ & 0 \\ \hline
$0.003{<}\ttod{<}0.4$   & 0.1899(7) & $-0.45(29)$ & $0.565(88)$ & 1.01(7) \\ \hline
$0.001{<}\ttod{<}0.2$  & 0.1879(7) & $-0.351(40)$ & $ 1.03(15)$ & 1.47(15) \\
                        & 0.1932(2) & $-0.718(24)$  & $  0.411(41)  $ & 0 \\ \hline
$0.003{<}\ttod{<}0.2$   & 0.1874(7) & $-0.321(39)$ & $1.31(15)$ & 1.55(14) \\
                        & 0.1945(9) & $-0.749(24)$ & $-0.461(40)$ & 0 \\ \hline
$0.001{<}\ttod{<}0.1$ & 0.1903(6) & $-0.587(22)$  & $0.122(45)$ & 0 \\
                        & 0.1889(3) & $-0.529(4)$  & $  0      $ & 0 \\ \hline
$0.003{<}\ttod{<}0.1$  & 0.1908(7) & $-0.605(23)$  & $0.156(46)$ & 0 \\
                        & 0.1888(3) & $-0.528(4)$  & $  0      $ & 0 \\ \hline
$0.001{<}\ttod{<}0.025$ & 0.1888(10) & $-0.482(72)$  & $0.19(23)$ & 0 \\
                        & 0.1895(4) & $-0.541(9)$  & $  0      $ & 0 \\ \hline
$0.003{<}\ttod{<}0.025$ & 0.1875(13) & $-0.398(82)$  & $0.44(25)$ & 0 \\
                        & 0.1897(4) & $-0.544(8)$  & $  0      $ & 0 \\ \hline
$0.001{<}\ttod{<}0.01$ & 0.1885(11) & $-0.511(30)$  & 0 & 0 \\
\hline\hline
\end{tabular}
\label{tab-c-eff-lt}
\end{table}

\begin{figure}[t]
\centering
\includegraphics[width=\columnwidth]{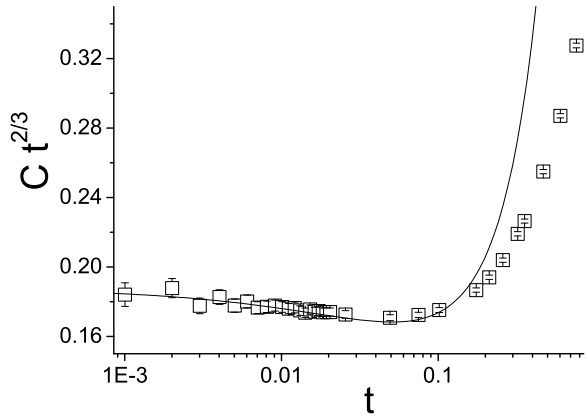}
\caption{Effective amplitude of the specific heat in the
high-temperature phase. Symbols show combined MC data for $L=243$ (for $t > 0.02$)
and $363$ (for $t \le 0.02)$,
and the solid line is the four-term approximation
Eqs.~(\ref{heat-joyce})--(\ref{heat-joyce-coeff}) to the exact
solution.} \label{c-eff-ht}
\end{figure}

\begin{figure}[t]
\centering
\includegraphics[width=\columnwidth]{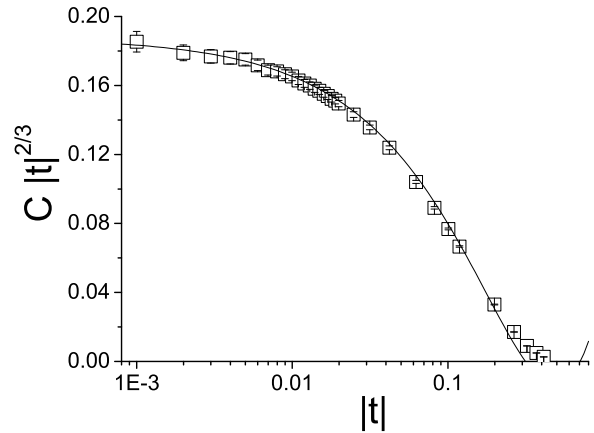}
\caption{Effective amplitude of the specific heat in the
low-temperature phase. Symbols and curves are the same as in
Fig.~\ref{c-eff-ht}.} \label{c-eff-lt}
\end{figure}

Figures~\ref{c-eff-ht} and \ref{c-eff-lt} show our MC data for the effective
specific-heat amplitude
in the high- and low-temperature phase, respectively, together with the
four-term approximation
Eqs.~(\ref{heat-joyce})--(\ref{heat-joyce-coeff}) to the exact solution,
which is seen to be in a fairly good agreement with the numerical data.

We fit the following expression for the effective amplitude of the specific heat
to the data:
\begin{equation}
A_{\rm eff}(\pm)(t)=A_0(\pm)+A_1(\pm)\ttod^\frac23+A_2(\pm)t +A_3(\pm)\ttod^\frac43
\label{fit-c-eff}
\end{equation}
by varying the temperature region window. Results of the fits are presented in
Tables~\ref{tab-c-eff-ht} and~\ref{tab-c-eff-lt}.
The data clearly support in both phases the specific-heat amplitude value
$A_0=0.1877\ldots$ rather well. The first correction-to-scaling amplitude
$A_1=-0.3884$ is estimated less well, but is still precise enough to give support
for the theoretically expected value $2/3$ of the correction-to-scaling exponent.
Higher correction-to-scaling terms could not be estimated from the available
MC data set.

\subsection{Energy}

The specific-heat amplitudes are connected with the energy amplitudes.
Let us define effective energy amplitudes as

\begin{eqnarray}
e_{\rm eff}(+)(t)&=&\left(e_+(t)-e_0\right) t^{-\frac13} \,, \nonumber \\
e_{\rm eff}(-)(t)&=&\left(e_-(t)-e_0\right) |t|^{-\frac13} \,.
\end{eqnarray}
In the vicinity of the critical point, they can be expanded as~\cite{Joyce-C}
\begin{equation}
e_{\rm eff}(\pm)(t)=E_1+E_2|t|^\frac23+E_3t+E_4|t|^{\frac43}+{\cal O}(|t|^\frac53)
\label{e-eff-p}
\end{equation}
\noindent with coefficients
\begin{eqnarray}
E_1&=& 1.278\,376\,401\ldots \,, \nonumber \\
E_2&=&-0.881\,371\,587\ldots \,,  \nonumber \\
E_3&=& 0.503\,377\,046  \nonumber \ldots \,, \\
E_4&=& 0.540\,143\,046\ldots \,.
\label{energy-coeff}
\end{eqnarray}

\begin{figure}[t]
\centering
\includegraphics[width=\columnwidth]{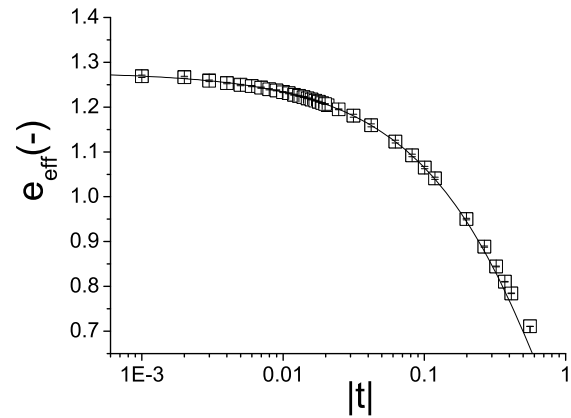}
\caption{Effective amplitude of the energy in the
low-temperature phase. Symbols are MC data and solid line is the
four-term series expansion~(\ref{e-eff-p}), (\ref{energy-coeff}).} \label{eff-e-lt}
\end{figure}

Figure~\ref{eff-e-lt} shows the effective amplitude of the energy in the
low-temperature phase together with the
four-term series expansion~(\ref{e-eff-p}), (\ref{energy-coeff}).
Fits to the energy amplitudes are shown in Table~\ref{tab-eeff-lt}, which
clearly support the first two coefficients in
(\ref{energy-coeff}). Fits to the effective energy amplitude in the
high-temperature phase look very similar.

\begin{table}[b]
\caption{Results of fits to the MC data for the effective energy
amplitude in the low-temperature phase.}
\center
\begin{tabular}{lllll}
\hline\hline
fit window & $E_1(-)$ & $E_2(-)$ & $E_3(-)$ & $E_4(-)$  \\ \hline
exact  & $1.2783\ldots$ &  $-0.8813\ldots$ &$0.5033\ldots$ &$0.5401\ldots$ \\ \hline
$0.001{<}\ttod{<}0.56$   & 1.2771(10) & $-0.80(4)$ & $0.86(8)$ & 0.99(5) \\
                     & 1.2946(4) & $-1.42(1)$ & $-0.68(1)$ & 0 \\ \hline
$0.001{<}\ttod{<}0.32$   & 1.2779(14) & $-0.84(6)$ & $0.86(8)$ & 0.99(5) \\
                         & 1.2834(8) & $-1.13(2)$ & $-0.68(1)$ & 0 \\ \hline
$0.001{<}\ttod{<}0.1$   & 1.2795(13) & $-0.99(4)$ & $0.01(9)$ & 0 \\
                         & 1.2796(6) & $-0.99(9)$ & $ 0$ & 0 \\ \hline
$0.001{<}\ttod{<}0.01$   & 1.2795(17) & $-0.99(5)$ & $0$ & 0 \\ \hline
$0.003{<}\ttod{<}0.01$   & 1.2779(23) & $-0.95(6)$ & $0$ & 0 \\ \hline\hline
\end{tabular}
\label{tab-eeff-lt}
\end{table}

\subsection{Magnetization}

From here on we follow the usual convention in the magnetic sector and use
the reduced temperature $\tau$ as independent variable in figures and fits.

\begin{figure}[b]
\centering
\includegraphics[width=\columnwidth]{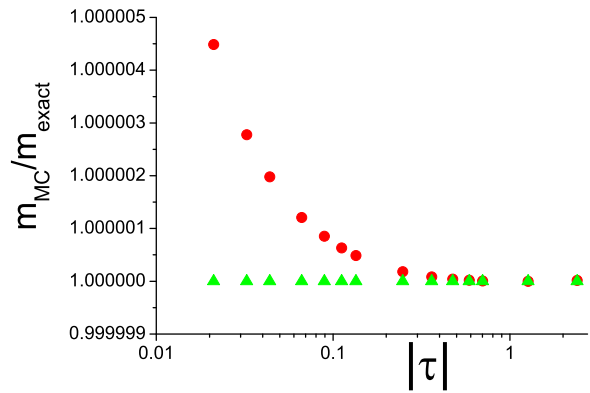}
\caption{Ratio of MC data with $L=363$ for the magnetization to the exact result. The magnetization $m_{\rm MC}$
is computed using Eqs.~(\ref{bw-mag}) (circles) and (\ref{bw-mag-alt}) (triangles).} \label{mag-rms-mod}
\end{figure}

The magnetization may be estimated using two methods, defined by Eqs.~(\ref{bw-mag})
and (\ref{bw-mag-alt}). Figure~\ref{mag-rms-mod} shows the ratio of the magnetization
to the exact value, computed using both methods. The relative difference reaches  $10^{-5}$
close to the left boundary of the critical region window. We also checked that the low-
and high-temperature susceptibilities are not very sensitive to the way the magnetization
is calculated and Fig.~\ref{mag-rms-mod} gives preference for using the
definition~(\ref{bw-mag-alt}) which we use in this paper for the calculation of the
magnetization and the magnetic susceptibility.  This was mentioned already in the
paper~\cite{Mark-David81}.

\begin{figure}[b]
\centering
\includegraphics[width=\columnwidth]{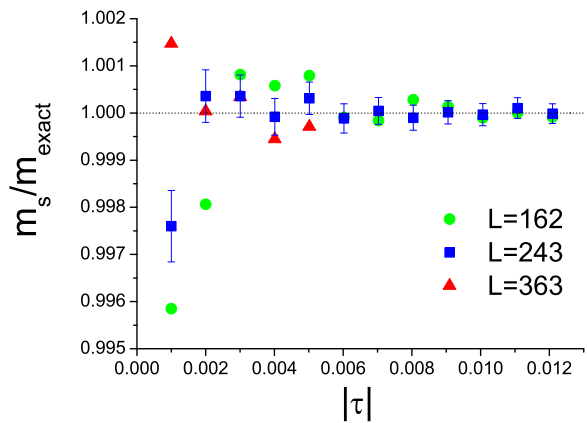}
\caption{Ratio of MC data for the magnetization $m_s$ to the exact result
$m_{\rm exact}$
for three values of lattice size. Error bars are only shown for the
lattice size $L=243$ (open squares).} \label{magnetization-ratio}
\end{figure}

Figure~\ref{magnetization-ratio} shows a comparison of MC data for the
magnetization with the expansion~(\ref{M-joyce-t}) of the exact result, where
the MC data for several lattice sizes are divided by the exact value.
We see that the MC data coincide rather well with the exact result down to
the reduced temperature $\tmod\approx 0.003$. This value thus defines the
left boundary of the critical region window in our subsequent analysis.

\begin{figure}[b]
\centering
\includegraphics[width=\columnwidth]{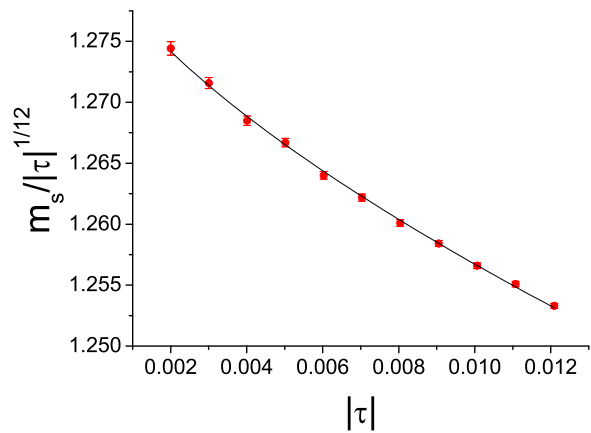}
\caption{Effective amplitude of the magnetization: Fit with expression~(\ref{ms-eff-fit}) (solid line)
to the data for $L=243$ (solid circles).} \label{mag-corr}
\end{figure}

The data for the effective amplitude $B_s=m_s/\tmod^{1/12}$ (see Fig.~\ref{mag-corr})
were fit with the expression
\begin{equation}
B_s=B_0+B_1\tmod^{2/3}+{\cal O}(\tmod) \label{ms-eff-fit} \,,
\end{equation}
which, up to this order, agrees with the expansion~(\ref{M-joyce-t}).

For $L=162$, in the temperature window $\tmod\in[0.003-0.012]$ (compare with Fig.~\ref{magnetization-ratio}),
the estimation gives an amplitude of $B_0=1.2842(4)$
which is three standard deviations off the exact value $B_0=1.28326\dots$. For $L=243$, estimated within
the appropriate temperature window $\tmod\in[0.002-0.012]$, the value of the critical amplitude
$B_0=1.2833(2)$ is in excellent agreement with the exact value. In both cases we found that in the
temperature windows it is sufficient to just include the first correction-to-scaling term and
we can neglect the second one in the fit. The estimated values of the first correction-to-scaling amplitude
are $B_1=-0.592(11)$ and $-0.582(6)$ for $L=162$ and $L=243$, respectively, which are in good agreement
with the exact value $B_1=-0.58982\dots$.  Figure~\ref{mag-corr} shows the fit to the effective amplitude
of magnetization data for $L=243$.

\begin{table}[b]
\caption{Results of fits to the MC data for the effective low-temperature susceptibility
amplitude $\Gamma_-^{\rm eff}$.}
\center
\begin{tabular}{l|lllll}
\hline\hline
fit window & $\Gamma_-$ & $D_1^-$ & $D_2^-$  &$D_3^-$\\ \hline
$0.002{<}\tmod{<}0.5$   & 0.06814(5) & $-0.423(3)$  & 0.491(12) &  $-0.11(1)$\\ \hline
$0.002{<}\tmod{<}0.25$   & 0.06751(6) & $-0.355(6)$  & 0.170(24) &  0.25(3) \\
                     & 0.06800(3) & $-0.406(2)$  & 0.401(3) &  0 \\ \hline
$0.003{<}\tmod{<}0.25$   & 0.06931(9) & $-0.468(7)$ & 0.592(29) & $-0.15(3)$  \\
                     & 0.06893(5) & $-0.435(2)$  & 0.449(3) &  0 \\ \hline
$0.002{<}\tmod{<}0.05$   & 0.06804(31) & $-0.308(38)$ & $-0.32(22)$ & 1.26(34)  \\
                     & 0.06909(10) & $-0.444(7)$  & 0.473(20) &  0 \\ \hline
$0.002{<}\tmod{<}0.025$  & 0.06655(8) & $-0.273(7)$ & $-0.042(23)$ & 0  \\
                     & 0.06669(3) & $-0.286(1)$ & 0 & 0  \\ \hline
$0.003{<}\tmod{<}0.025$  & 0.06879(13) & $-0.420(9)$ & 0.39(3) & 0  \\
                     & 0.06715(4) & $-0.295(1)$ & 0 & 0  \\ \hline
$0.003{<}\tmod{<}0.011$  & 0.06770(6) & $-0.313(2)$ & 0 & 0  \\ \hline
$0.004{<}\tmod{<}0.011$  & 0.06801(9) & $-0.32(3)$ & 0 & 0  \\
\hline\hline
\end{tabular}
\vspace*{0.5cm}
\label{tab-chi-lt}
\end{table}

We also checked the equidistribution of magnetization moments $\langle m^n_i\rangle$ and polarization moments $\langle p^n_i\rangle$
with $n=1,2,4$ over the three sub-lattices $L_i$ and $L_{jk}$ and found it valid within statistical accuracy.

\subsection{Polarization}

The critical amplitude $P_0$ of the polarization
$$ P=P_0\; \tmod^{1/12}+ \ldots $$
can be estimated in the same manner as the magnetization
critical amplitude $B_0$. The final value is $P_0=1.2104(3)$, to be compared with the exact value~\cite{LT-exact}
$$ P_0=2^\frac{15}{8} \left(\ln(\sqrt{2}+1)\right)^\frac{1}{12}\!\!/3 \approx 1.20987\dots \; .$$
We may also estimate the ratio $B_0/P_0=1.061(1)$ which is close to the apparently exact
value $3\sqrt{2}/4\approx 1.06066$.

\subsection{Low-temperature susceptibility}

Let us now come to the main subject of the present paper, the
magnetic susceptibility amplitudes which are not known analytically. We
can estimate them from our MC data using the same type of analysis
we performed for the specific heat in the previous subsection. For
an additional control of the analysis we compare our MC data with the available
series expansions (SE) data. The available SE data~\cite{LT-SE} are short, however, and could
not be used for a reliable estimation of amplitude values.

\begin{figure}[tb]
\centering
\includegraphics[angle=0,width=\columnwidth]{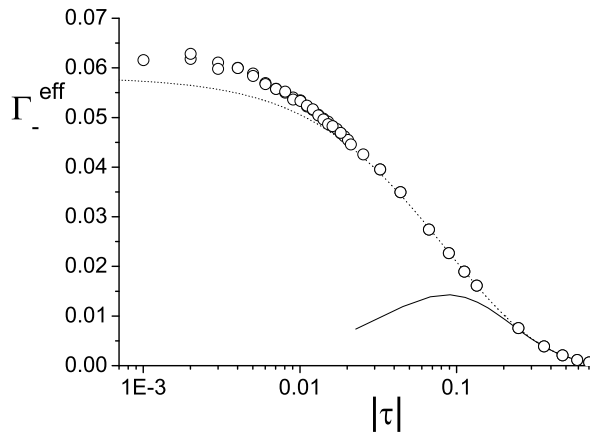}
\caption{Effective amplitude $\Gamma_-^{\rm eff}$ of the
magnetic susceptibility in the low-temperature phase: MC data (circles),
SE data (solid line), Pad\'e approximant to SE data (dashed line).} \label{chi-lt-se}
\end{figure}

Figure~\ref{chi-lt-se} shows MC and SE data for the effective amplitude susceptibility
$\Gamma^{\rm eff}_-=k_BT\chi_- \tmod^{7/6}$ in the low-temperature phase as function
of $-\tau$. The solid
line represents a direct SE data summation, while the dashed line is
the Pad\'e approximant of SE data~\cite{Paolo-private}. MC and SE data coincide well for
$\tmod>0.01$, and the discrepancy for smaller $\tmod$ is not
surprising due to the short SE data.

The parameters of the fit to the MC data shown in Fig.~\ref{chi-lt-se}
according to the expression
\begin{equation}
\Gamma^{\rm eff}_-=\Gamma_-+D_1^-\tmod^\frac23+D_2^-\tmod+D_3^-\tmod^\frac43
\label{fit-x-eff}
\end{equation}
are given in Table~\ref{tab-chi-lt}. Clearly, we can accept as the final and very
conservative estimate the value $\Gamma_-=0.0681(1)$.

Using the exact values for $A_0$ ($= A_0(-)=A_0(+)$)  and $B_0$, we can estimate
from this value the universal ratio
\begin{equation}
R_C^- = \alpha A_0 \Gamma_-/B_0^2 = 0.00517(7) \,.
\end{equation}

We also estimated the low-temperature critical amplitude of the polarization
susceptibility $\Gamma_-^{(p)}$ in the same manner as for the magnetic
susceptibility. The result is $\Gamma_-^{(p)}=0.061(1)$ and
$\Gamma_-/\Gamma_-^{(p)}\approx 1.11$ which appears to be close to the ratio
$(B_0/P_0)^2 = 1.12499\dots$.

\subsection{High-temperature susceptibility}

\begin{figure}[tb]
\centering
\includegraphics[angle=0,width=\columnwidth]{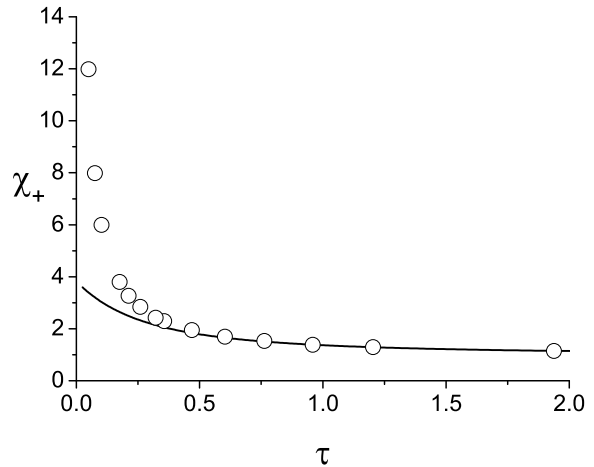}
\caption{Magnetic susceptibility $\chi_+$ in the high-temperature phase: MC
data (circles), SE data (line).} \label{chi-ht-se}
\end{figure}

Figure~\ref{chi-ht-se} shows MC and SE~\cite{HT-SE} data for the susceptibility
$\chi_+$ in the high-temperature phase as function of $\tau$. The data
coincide well at large enough reduced temperature $\tau<0.5$ and
diverge at small $\tau$ because of the small number of terms in the
SE available.

\begin{figure}[tb]
\centering
\includegraphics[angle=0,width=\columnwidth]{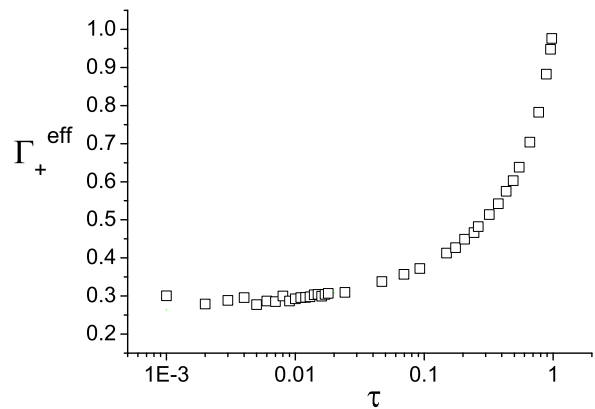}
\caption{Effective amplitude of the magnetic susceptibility in the
high-temperature phase. } \label{chi-ht-eff}
\end{figure}

The effective amplitude of the high-temperature susceptibility is shown in
Fig.~\ref{chi-ht-eff}. Our results of fits of the effective amplitude data
with the expression
\begin{equation}
\Gamma^{\rm eff}_+=\Gamma_++D_1^+\tau^\frac23+D_2^+\tau+D_3^+\tau^\frac43
\end{equation}
are presented in Table~\ref{tab-chi-ht}.

\begin{table}[tb]
\caption{Results of fits to the MC data for the effective high-temperature susceptibility
amplitude $\Gamma_+^{\rm eff}$.}
\center
\begin{tabular}{l|lllll}
\hline\hline
fit window & $\Gamma_+$ & $D_1^+$ & $D_2^+$  &$D_3^+$\\ \hline
$0.002{<}\tmod{<}0.77$   & 0.265(2) & $0.79(7)$  & $-0.93(18)$ &  0.79(12)\\
                         & 0.276(1) & $0.34(2)$  & 0.25(2) &  0 \\ \hline
$0.002{<}\tmod{<}0.43$    & 0.267(3) & $0.68(12)$  & $-0.59(34)$ &  0.53(27)\\
                         & 0.272(1) & $0.45(3)$  & 0.09(4) &  0 \\ \hline
$0.002{<}\tmod{<}0.025$   & 0.266(3) & $0.74(48)$  & $-0.7(1.5)$ &  0\\
                         & 0.265(1) & $0.549(3)$  & 0 &  0 \\ \hline
$0.004{<}\tmod{<}0.025$   & 0.270(3) & $0.52(6)$  &  0&  0\\ \hline
$0.002{<}\tmod{<}0.01$   & 0.268(4) & $0.57(12)$  &  0&  0\\ \hline
$0.004{<}\tmod{<}0.01$   & 0.269(7) & $0.54(18)$  & 0&  0\\
\hline\hline
\end{tabular}
\vspace*{0.5cm}
\label{tab-chi-ht}
\end{table}

The final estimate of the high-temperature susceptibility amplitude is
$\Gamma_+=0.265(5)$. This implies for the universal susceptibility
amplitude ratio the central estimate
\begin{equation}
{\cal R}_\chi \equiv \Gamma_+/\Gamma_-=3.9(1) \,,
\end{equation}
in very good agreement with the analytical predictions of Refs.~\cite{Delfino98}
and \cite{DG04}.

\section{Discussion}
\label{sec-discussion}

It is a widely accepted believe that there are four known models in the
4-state Potts model universality class. Besides the 4-state Potts model
itself~\cite{Potts}, these are the Baxter-Wu model~\cite{BW}, the
Ashkin-Teller model with some particular values of parameters~\cite{Baxter},
and the Debierre-Turban model~\cite{DT83} with some particular value of parameter.
Table~\ref{tab-summary} summarizes the known knowledge of universal amplitude
ratios for the first three models, where we also included the universal ratio
 $R_C^+ = \alpha A_0(+) \Gamma_+/B_0^2$. There are no estimations made for the
Debierre-Turban model. Clearly, all estimates for $\Gamma_+/\Gamma_-$ from
Monte Carlo simulations and series expansions are systematically smaller
than the analytical predictions, also compatible within error bars, with
a higher deviation reported in \cite{CasTatVin99}. There are some visible
deviations of the result published in \cite{Shchur09}\footnote{We mistakenly
mentioned in~\cite{bbjs} a wrong estimation of the value of $\Gamma_+/\Gamma_-$
for the Baxter-Wu model.}. The analysis of this quantity presented in
Ref.~\cite{Shchur09} is based on the inclusion of logarithmic corrections
to scaling (both multiplicative and additive) in the fit. This procedure is
a bit risky although one does not have to do something else. At the same
time, it is argued in \cite{Shchur09} that the universal combination $R_C^-$
should not contain any logarithmic corrections in the effective estimation
through the function $R_C(\tmod)^-=\alpha(\alpha-1)(e_-(\tmod)-e_0)\chi_-/m^2(\tmod)$
and, indeed, there is a good coincidence of results reported in
Refs.~\cite{Delfino98,Shchur09}, and in the present paper for $R_C^-$.
By analyzing the data in Table~\ref{tab-summary}, we may conclude
that there are definite overestimations of the critical amplitude
$\Gamma_+$ in Ref.~\cite{Shchur09}. This is possibly due to the large
background terms (nonsingular contribution) in the high-temperature susceptibility.
We have to note that there is only one direct estimate of the universal
ratio $\Gamma_T/\Gamma_-$ published in \cite{Shchur09}, which is not
consistent with analytical predictions. More work should be done to clarify this
issue.

Finally, we may conclude that our analysis of critical amplitudes of the
Baxter-Wu model produces universal amplitude ratios consistent with the
analytical results obtained by Cardy and Delfino for the 4-state Potts
model~\cite{Delfino98} and by Delfino and Grinza for the special case of
the Ashkin-Teller model~\cite{DG04}.

\begin{table}[tb]
\caption{Universal combinations of critical amplitudes for the
two-dimensional models in the 4-state Potts model universality class.}
\center
\begin{tabular}{l|l|lllll|l}
\hline\hline
model & approach & $A_+/A_-$ & $\Gamma_+/\Gamma_-$ &$\Gamma_T/\Gamma_-$ & $R_C^-$ & $R_C^+$ & Ref. \\ \hline
4-state Potts model & analytical & 1.0      & 4.013    & 0.129     & 0.00508    & 0.0204    & \cite{DelfinoBarkemaCardy00,Delfino98} \\
                    & MC         & $-$      & 3.14(70) & $-$       & 0.0068(9)  & 0.021(5)  & \cite{CasTatVin99} \\
                    & SE         & $-$      & 3.5(4)   & $-$       & $-$        &  $-$      & \cite{Enting03} \\
                    & MC and SE  & 1.000(5) & 6.49(44) & 0.154(12) & 0.0052(2)  & 0.0338(9) & \cite{Shchur09} \\ \hline
Ashkin-Teller model & analytical & $-$      & 4.02     & 0.129     & $-$        &  $-$      & \cite{DG04} \\ \hline
Baxter-Wu model     & MC         & 0.995(5) & 3.9(1)   & $-$       & 0.00517(7) & 0.0201(5) & present \\
\hline\hline
\end{tabular}
\vspace*{0.5cm}
\label{tab-summary}
\end{table}

\acknowledgments The authors wish to thank Paolo Butera who kindly computed
Pad\'e approximants for susceptibility series expansions. We appreciate useful
discussions with B. Berche, P. Butera and F. Igloi. This work is supported
by the Deutsche Forschungsgemeinschaft (DFG) through Grant-No.\ 436
RUS 17/122/03 and by the Russian Foundation for Basic Research. WJ gratefully
acknowledges support by the Research Academy Leipzig (RAL) and the
top-level research area PbF2 ``Mathematical Sciences'' of the University
of Leipzig.

\end{document}